\documentclass[useAMS,usenatbib]{mn2e}
\usepackage{amssymb}
\usepackage{amsmath}
\usepackage{txfonts}
\usepackage{graphicx}
\usepackage{natbib}
\usepackage{rotating}
\usepackage{url}
\usepackage{epsfig}
 
\newif\ifpdf\ifx\pdfoutput\undefined\pdffalse\else\pdfoutput=1\pdftrue\fi

\def\bi {\begin{itemize}}
\def\ei {\end{itemize}}

\def\intgr{\textit {INTEGRAL}}

\def\rxte{\textit {RXTE}}
\def\asca{\textit {ASCA}}
\def\ascasp{\textit {ASCA }}
\def\xmm{\textit {XMM-Newton}}
\def\xmmsp{\textit {XMM-Newton }}
\def\psrb{PSR~B1259--63} 
\def\psrbsp{PSR~B1259--63 } 

\def\sax{\textit{Beppo}SAX}
\def\rxp {2RXP~J130159.6-635806}
\def\rxpsp {2RXP~J130159.6-635806 }

\usepackage{color}
\definecolor{red}{rgb}{0.7,0,0}
\definecolor{blue}{rgb}{0,0,0.7}

\begin{document}

\title{\xmm\ observations of \psrb{} near the 2004 periastron passage} 

 \author[M. Chernyakova et al.]{M. Chernyakova$^{1,2}$\thanks{E-mail:Masha.Chernyakova@obs.unige.ch}\thanks{M.Chernyakova 
is on leave from Astro Space Center of the P.N.~Lebedev Physical
Institute,  Moscow, Russia}, A. Neronov$^{1,2}$, A. Lutovinov$^{3}$,
 J. Rodriguez$^{4}$ and S. Johnston$^{5}$
 \\
$^{1}$INTEGRAL Science Data Center, Chemin d'\'Ecogia 16, 
1290 Versoix, Switzerland\\
$^{2}$Geneva Observatory, 51 ch. des Maillettes,
CH-1290 Sauverny, Switzerland \\
$^{3}$ Space Research Institute,  84/32 Profsoyuznaya Street, 
Moscow 117997, Russia\\
$^{4}$ CEA Saclay, DSM/DAPNIA/Service d'Astrophysique (CNRS UMR 7158 AIM),
91191 Gif sur Yvette, France\\
$^{5}$ Australia Telescope National Facility, CSIRO, P.O. Box 76,
Epping, NSW 1710, Australia.}

\date{Received $<$date$>$  ; in original form  $<$date$>$ }
\pagerange{\pageref{firstpage}--\pageref{lastpage}} \pubyear{2005}

\maketitle
\label{firstpage}

\begin{abstract}
  \psrb{} is in a highly eccentric 3.4 year orbit with a Be star and
  crosses the Be star disc twice per orbit, just prior to and just
  after periastron. Unpulsed radio, X-ray and gamma-ray emission
  observed from the binary system is thought to be due to the
  collision of pulsar wind with the wind of Be star.  We present here
  the results of new \xmm\ observations of the \psrbsp system during
  the beginning of 2004 as the pulsar approached the disc of Be star.
  We combine these results with earlier unpublished X-ray data from
  \sax\ and \xmm\ as well as with \asca\ data. The detailed X-ray
  lightcurve of the system shows that the pulsar passes (twice per
  orbit) through a well-defined gaussian-profile disk with the
  half-opening angle (projected on the pulsar orbit plane)
  $\Delta\theta_{disk}\simeq 18.5^\circ$. The intersection of the disk
  middle plane with the pulsar orbital plane is inclined at
  $\theta_{disk}\simeq 70^\circ$ to the major axis of the pulsar
  orbit.  Comparing the X-ray lightcurve to the TeV lightcurve of the
  the system we find that the increase of the TeV flux some 10--100
  days after the periastron passage is unambiguously related to the
  disk passage.  At the moment of entrance to the disk the X-ray
  photon index hardens from $\Gamma\simeq 1.8$ up to $\Gamma\simeq
  1.2$ before returning to the steeper value $\Gamma\ge 1.5$. Such
  behaviour is not easily accounted for by the model in
  which the X-ray emission is synchrotron emission from the
  shocked pulsar wind. We argue that the observed hardening of the
  X-ray spectrum is due to the inverse Compton or bremsstrahlung
  emission from 10-100 MeV electrons responsible for the radio
  synchrotron emission.
\end{abstract}


\section{Introduction}

\psrb{} is a $\sim$48 ms radio pulsar in a highly eccentric
(e$\sim$0.87), 3.4 year orbit with a Be star SS 2883
\citep{johnston92}.  
The system's distance from Earth is $d\sim 2$ kpc \citep{johnston94},
note, however, that $d$ is uncertain by a factor of $\sim$2 based on radio 
measurements (see, e.g., Manchester, Taylor, \& Lyne 1993).

Be stars are well-known to be sources of strong
highly anisotropic matter outflow. Both a dilute polar wind and a
denser equatorial disc have been invoked to reconcile models for
infra-red, ultra-violet and optical observations \citep{waters88}.
Timing analysis of the \psrb{} system shows that the disc of Be star
is tilted with respect to the orbital plane \citep{wex98,wang04}.
The properties of the radio emission from the system are very
different close to and far from the periastron.  Radio observations of
the 1994, 1997 \citep{johnston99}, 2000 \citep{connors02}, and the
2004 \citep{johnston05} periastron passages show that when the pulsar
is far from periastron the observed radio emission is comprised
entirely in highly linearly polarized pulsed emission from the pulsar
itself with an intensity practically independent of the pulsar orbital
position \citep{McClure98}.  But, starting about 100 days before
periastron, depolarization of the pulsed emission occurs, the
dispersion measure and the absolute value of the rotation measure
increase while the flux density decreases. The pulsed radio emission
then disappears entirely as the pulsar enters the disc and is hidden
behind it (relative to the observer) approximately 20 days before the
periastron passage.  Shortly before the disc crossing unpulsed radio
emission appears and within several days sharply rises to a peak which
is more than ten times higher than the intensity of the pulsed
emission far from the periastron.  Afterward the unpulsed flux
slightly decreases, as the pulsar passes through periastron before
reaching a second peak just after the pulsar crosses the disc for the
second time.  The unpulsed radio emission is detected until at least
100 days after the periastron passage.
\citep{johnston99,johnston05,connors02}.

The first detection of the X-rays from \psrb\ system was done by
 \textit{ROSAT} in 1992 \citep{cominsky94}.
The most recent published observations of the system in X-rays were
carried out with the \ascasp satellite in 1994 and 1995
\citep{kaspi95,hirayama99}. These observations show that X-ray
emission is approximately twice as high at the time of the disc
crossing than at periastron. In soft gamma-rays the system was
observed only after the periastron passage with CGRO/OSSE and
\intgr/ISGRI \citep{grove95,shaw04}.  The combined soft gamma- and
X-ray spectrum is consistent with a power law of photon index $\sim
1.7$.  This index also coincides with the radio spectral index of the
unpulsed emission. No pulsed X-ray emission was detected from the
system. With the standard epoch-folding method the 
upper limit to a pulsed component in 2 -- 10 Kev energy
 range was estimated to be about 10\% of the unpulsed component 
close to the periastron, and to about 40\% during the apastron
\citep{kaspi95,hirayama99}. Note that search for the pulsations in
P - \.P space allowed \citet{kaspi95} to reduce an upper limit 
close to the periastron to less than 2\% under a certain assumptions 
on the shape of pulse profile.

The 2004 periastron passage was observed in TeV
gamma-rays by HESS \citep{aharon05}. TeV lightcurve of the source
shows an unexpected behaviour. Contrary to the prediction that the TeV
flux should be maximum at the periastron \citep{kirk02},
 the flux apparently has a
local minimum around this moment.

Several theoretical models were put forward to explain the behaviour
of the X-ray lightcurve of the system. In the model of
\citet{tavani97}, the enhancement of {\it synchrotron} X-ray emission
during the disk passage is attributed to enhancement of magnetic field
at the position of the contact surface of pulsar and stellar winds. In
the model of \citet{cher99,cher00} the enhancement of the {\it inverse
  Compton} X-ray emission during the disk passage is supposed to be
due to the effect of the macroscopic mixing of the stellar and pulsar
winds in the disk which effectively enhances the escape time of
high-energy electrons responsible for the observed emission.

Typical energies of electrons responsible for the X-ray emission are
$E_e\sim$~TeV in \citet{tavani97} model while in \citet{cher99} model
$E_e\sim 10$~MeV.  One can, in principle, distinguish between the two
models by comparing the details of radio, X-ray and TeV lightcurve of
the system.  We have organized a cycle of five \xmmsp observations as
the pulsar approached the 2004 periastron. In this paper we present
the results of our observations along with other \xmm\ observations of
the system not previously published. For the completeness we also
include the analysis of the previously unpublished \sax\ data. The
2004 periastron passage was extensively observed in radio
\citep{johnston05} and TeV \citep{aharon05} bands. Combination of
the radio, X-ray and TeV data opens up a possibility, to study the
evolution of the spectrum of \psrb\ system over some 12 decades in
energy.

Our analysis of large X-ray data set enables, for the first time, to
constrain the geometrical parameters of the disk of Be star, such as
the opening angle and orientation. Comparing the TeV and X-ray
lightcurves we find that the increase of the TeV flux some 10 to 100
days after the periastron passage  is unambiguously related to
the pulsar passage through the Be star disk for the second time. This
challenges the conventional interpretation of the observed TeV
emission as the inverse Compton emission from the pulsar wind
electrons.

Another surprising result of our analysis that is not predicted 
in the theroetical models cited above is the persistently hard
X-ray spectrum (photon index $\Gamma\simeq 1.2-1.3$) observed during
the 10-day period when the pulsar enters the disk. 

This paper is organized as follows: in Section 2 we describe the
details of the \xmmsp and \sax\ data analysis.  The results are
presented in Section 3, and discussed in Section 4.

\section[]{Observations and Data Analysis}

\begin{figure}
\includegraphics[width=8.5cm,angle=0]{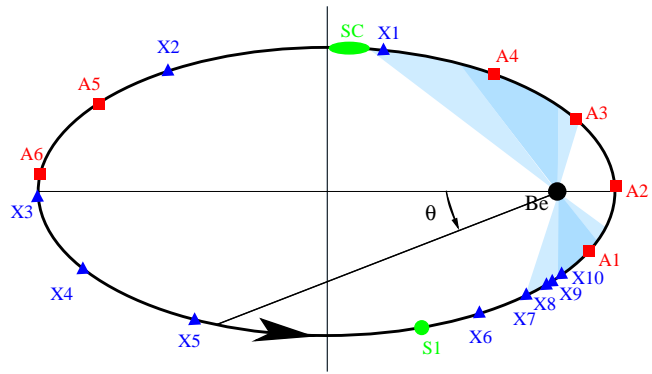}
\caption{Schematic representation of  PSR B1259-63 binary system with the locations of
  the pulsar during \xmm\, \sax\, and \textit{ASCA} observations.
  Shaded area shows the geometry of the disk inferred from the X-ray
  data. Darker and lighter shading corresponds to 1 and 2 half-opening
  angles of the disk, respectively. Angle $\theta$ is the orbital
  phase of the pulsar referred in the text.}
\label{xmm_ellips}
\end{figure}

In Figure \ref{xmm_ellips} we show a schematic drawing of the pulsar
orbit along with the approximate location of the pulsar position
during \xmm, \sax, and \asca\ observations. The X-ray observations are
more-or-less homogeneously distributed over the whole range of pulsar
orbital phase $0\le\theta<360^\circ$. The figure also shows the
location of the Be star disk (parameters of the disk are found below
from the analysis of the X-ray data). One can see that the set of
observations denoted as X6 -- X10 enables to study the entrance to the
disk in great details.

\subsection{\xmm\ observations}

The log of the \xmm\ data analyzed in this paper is presented in Table
1.  Observations X6 -- X10 were obtained as part of the 2004
periastron campaign and we have also used all other \xmm\ observations
of the system, which are now public (X1 -- X5).  In the Table negative
$\tau$ denotes the number of days before the date of 2004 periastron
passage (2004 March 7), and positive values correspond to the number
of days after the 2000 periastron (2000 October 17). $\theta$ is the
orbital phase counted as shown in Fig. \ref{xmm_ellips}.
\begin{table}
\caption{Journal of \xmm\ observations of \psrb \label{data}}
\begin{tabular}{c|c|c|c|c|c}
\hline
Data& Date & MJD& $\tau$&$\theta$&Exposure \\
 Set&      &    &  (days)&(deg)&(ks)         \\
\hline
X1&  2001-01-12 & 51921.73 & 87.2    & 320.8 &  11.3 \\
X2&  2001-07-11 & 52101.31 & 266.7   & 342.9 &   11.6 \\
X3&  2002-07-11 & 52467.24 & $-$604.1& 0.5   &  41.0 \\
X4&  2003-01-29 & 52668.27 & $-$403.1& 9.31  & 11.0 \\
X5&  2003-07-17 & 52837.53 & $-$233.9& 19.5  &  11.0 \\
X6&  2004-01-24 & 53028.79 & $-$42.6 & 57.4  &   9.7 \\
X7&  2004-02-10 & 53045.43 & $-$25.0 & 73.4  &   5.2  \\
X8&  2004-02-16 & 53051.39 & $-$20.0 & 83.1  &    7.7 \\
X9&  2004-02-18 & 53052.02 & $-$18.4 & 86.5  &   5.2 \\
X10& 2004-02-20 & 53055.82 & $-$15.6 & 93.2  &    6.9 \\				  
\hline
\end{tabular}
\end{table}

The \xmmsp Observation Data Files (ODFs) were obtained from the on-line
Science
Archive\footnote{http://xmm.vilspa.esa.es/external/xmm\_data\_acc/xsa/index.shtml};
the data were then processed and the event-lists filtered using {\sc
  xmmselect} within the Science Analysis Software ({\sc sas}) v6.0.1.
In all observations the source was observed with the MOS1 and MOS2
detectors.  The X1 -- X5 observations were done in the Full Frame
Mode, while the 2004 observations were performed in the Small Window
Mode, to minimize the pile-up problems. For the X6 -- X10 observations
PN data are also available.  In all observations a medium filter was
used.

The event lists for spectral analysis were extracted from a
45$^\prime$$^\prime$ radius circle at the source position for the X1
-- X5 observations, and from a 22.5$^\prime$$^\prime$ radius circle
for MOS 1,2 observations in Small Window Mode (X6 -- X10). For the PN
instrument a region of 35$^\prime$$^\prime$ around the source position
was chosen.  Background photons were collected from a region located
in the vicinity of the source with the same size area as the one
chosen for the source photons.

For the spectral analysis, periods of soft proton flares need to be
filtered out. To exclude them we have built light curves with a
hundred second binning and excluded all time bins in which the count
rate above 10 keV was higher than 1.5 cnt/s.   The source countrate in 1 
-- 10 keV energy range  varied from about 0.07  to 1.2 cts/s for MOS 
instruments, and from 0.4 to 4 cts/s for PN.  
Data from MOS1, MOS2
and, when available, PN detectors were combined in the spectral
analysis to achieve best statistics.
 

\subsection{\sax\ observations}
Table \ref{saxobs} lists the observations of \psrb{} made with \sax\ 
around the 1997 periastron passage.  In the Table $\tau$ denotes the
number of days before or after the 1997 periastron passage.
 
We used data of MECS instruments aboard BeppoSAX observatory which
provides the coverage in the energy range $\sim$1.5-11.0 keV.
Reduction of MECS data was done with the help of standard tasks of
LHEASOFT/FTOOLS 5.2 package according to recommendations of BeppoSAX
Guest Observer
Facility\footnote{http://heasarc.gsfc.nasa.gov/docs/sax/abc/saxabc/saxabc.html,
 http://www.asdc.asi.it/bepposax/software/cookbook/}. Spectrum of
the source was obtained extracting events from a circle of $3^\prime$
radius around the source position. Spectrum of the instrument
background was estimated considering events from a circle of the same
radius $6^\prime$ away from the source. Typically the source provided
several hundreds of counts per observation. Response matrices (RMF
and ARF) of MECS detectors were taken from HEASARC
archive\footnote{http://heasarc.gsfc.nasa.gov/FTP/sax/cal/responses}.
Corrections of photons arrival time to the Solar system barycenter for
subsequent usage of the event lists in timing analysis were done with
the help of tasks from SAXDAS
package\footnote{http://www.asdc.asi.it/bepposax/software/saxdas/index.html}.

\begin{table}
\caption{Journal of \sax\ observations of \psrb \label{saxobs}}
\begin{tabular}{c|c|c|c|c|c}
\hline
Data& Date & MJD& $\tau$&$\theta$&Exposure \\
 Set&      &    &  (days)&(deg)&(ks)         \\
\hline
S1& 1997-03-22  & 50529.63 & $-$68.31 & 44.9& 28.622\\
S2& 1997-09-02  & 50693.75 &  95.81 &  322.9& 16.985\\
S3& 1997-09-08  & 50699.44 & 101.50 &  324.1& 53.555\\
S4& 1997-09-17  & 50708.08 & 110.14 &  325.9& 51.301\\
S5& 1997-09-25  & 50716.05 & 118.11 &  327.3& 28.719\\
\hline
\end{tabular}
\end{table}
We have checked that all the spectra obtained in the S2 -- S5 observations
are consistent with each other, and combined them in order to
improve statistics. This combined spectrum we denote as SC hereafter.

\section{Results}

\subsection{Imaging Analysis}
\begin{figure}
\begin{center}
\includegraphics[width=8cm,angle=0,bb=35 155 510 640]{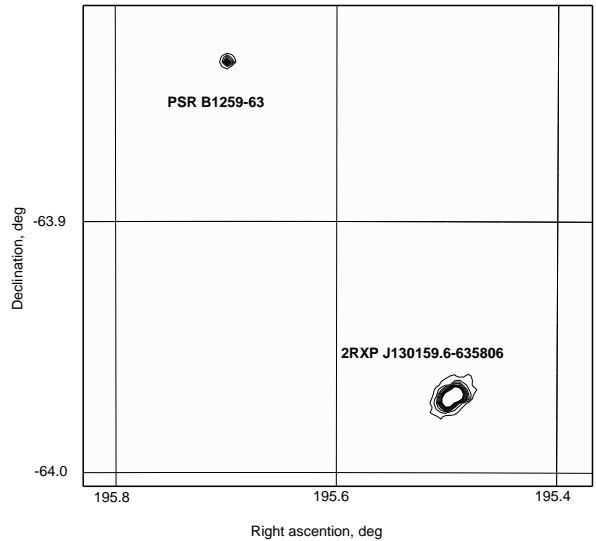}
\end{center}
\caption{Contour plot of the \xmm\ field of view for the X6 observation. A
total of 10 contours were used with a linear scale between 
5 counts per pixel (outer contour) and 50 counts per pixel (inner
contour). In this observation \rxp\ was forty times
brighter than \psrb. This figure is taken from Chernyakova et al. (2005).}
\label{xmm_ima}
\end{figure}
\begin{figure*}
\begin{center}
\includegraphics[width=14cm,angle=0]{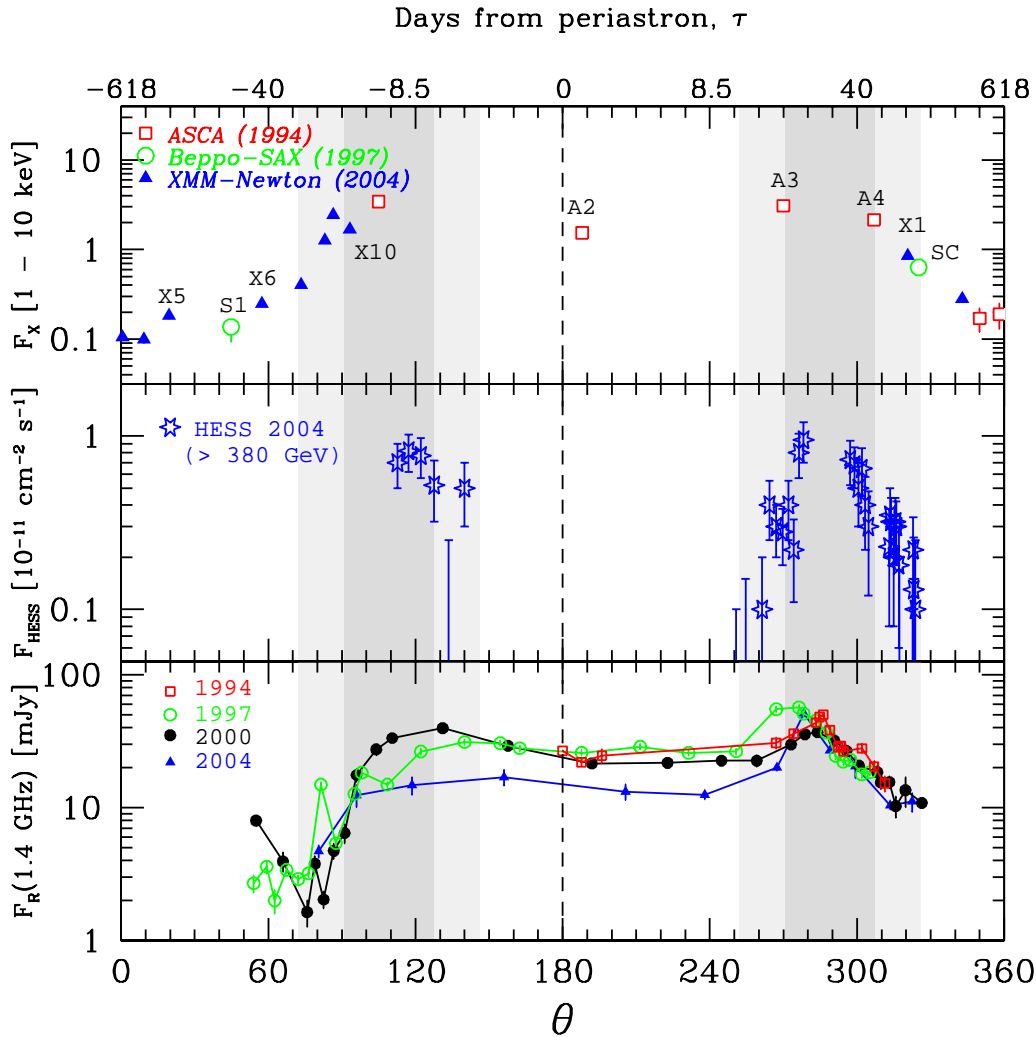}
\end{center}
\caption{Comparison between the X-ray (top), TeV (middle) and radio
  (bottom) lightcurves. The 1 - 10 keV X-ray flux is given in
  $10^{-11}  \mbox{ergs cm}^{-2} \mbox{s}^{-1}$.
  \xmm\ observations are marked with triangles,
  \sax\ ones with circles, and \asca\ ones with squares. Data for four
  different periastron passages are shown with different colors: red
  (1994), green (1997), black (2000) and blue (2004). Bottom X axis
  shows the orbital phase, $\theta$, top X axis shows days from
  periastron, $\tau$.}
\label{Xradio}
\end{figure*}

In the \xmm\ field of view during its \psrb\ observation program, two
sources were clearly detected (see Fig. \ref{xmm_ima} for the contour
plot of \xmm\ field of view for the MJD 53029 (X6) observation).
Besides \psrb\ itself there is  second source, which we have identified
as \rxp. Detailed analysis of this source are presented in
\citet{cher05}. In this paper it is shown in particular,
that \rxpsp is highly variable, with an intensity 
during flares which is several times larger than the peak intensity of
\psrb.
Therefore, although many observations of \psrb{} have been made with
\rxte, the fact that it is a non-imaging telescope means that it
is extremely difficult to determine the flux from \psrb{} alone.
We therefore do not use  numerous
\rxte\ observations of the 2004 periastron passage, presented in  
\citet{shaw04}. Note that despite in the \citet{shaw04} 
all \rxte\ observations are
 analysed, the authors were able to use in their
analysis only data simultaneous to \intgr\ observations, when it was clear
 that influence of \rxp\ is small.

\subsection{The X-ray lightcurve}

The upper panel of Fig. \ref{Xradio} shows the X-ray lightcurve of the
system. For comparison we also show in the same Figure TeV lightcurve
of 2004 HESS observation \citep{aharon05} and radio
\citep{johnston99,connors02,johnston05} lightcurves of different
years periastron passages.

The X-ray flux from the source, in agreement with previous observations,
was observed to be highly variable. The 1 -- 10 keV flux is at 
minimum near apastron (observation X3), and has  maximum more than
20 times larger, 18 days before periastron (X9).
During observations X5, X6, and X7, the intensity of the 
emission increased, as the pulsar 
approached periastron and the Be star disc.
Five days after X7 observation the flux increased by 
a factor of 4.
During the next two days the intensity continued
to increase, and reached its peak value at X9.
The next observation  (X10) was done only three days later, and the
observed flux was already one and a half times lower.

Comparison of X-ray and radio lightcurves shows that the X-ray flux
from the system varies in a more regular way than the radio flux: the
X-ray data points from different periastron passages lie more-or-less on the
same curve. Rapid grows of the X-ray flux found in our \xmm\
observations X6 -- X10 is correlated with the rapid growth of the 
unpulsed radio emission from the system. The growth of radio
and X-ray flux at these phases can be attributed to the pulsar
entering the Be star disk.  

Unfortunately, TeV observations start somewhat later and it is not
possible to see whether the TeV flux growth during the pre-periastron
disk crossing. However, simple geometrical argument tells that the
orbital phase $\theta$ at which the pulsar should enter the disk for
the second time should be shifted by $180^\circ$ relative to the first
entrance.  From Fig.  \ref{Xradio} one can infer that the first
pre-periastron entrance falls roughly between the phases
$70^\circ<\theta<110^\circ$. Thus, the pulsar has to enter the disk
again between the phases $250^\circ<\theta<290^\circ$. Surprisingly,
one can clearly see from the middle panel of Fig. \ref{Xradio} that
the TeV flux grows in this phase interval. 
\begin{figure}
\includegraphics[width=8cm,angle=0]{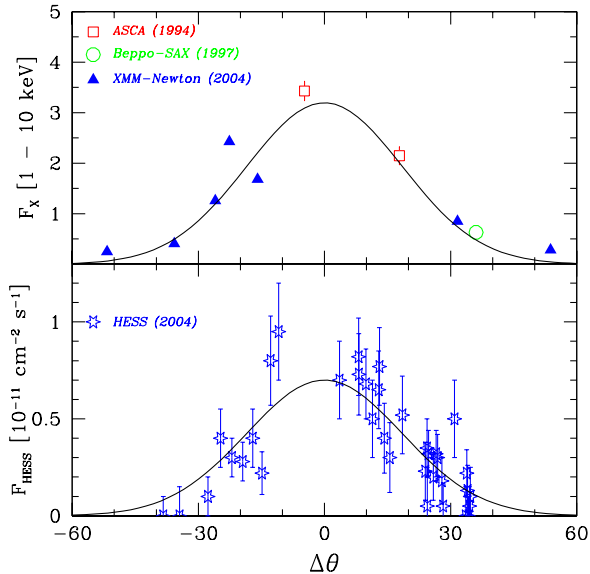}
\caption{The X-ray (top) and TeV (bottom) flux as a function of the relative
  phase $\Delta\theta=\theta-\theta_{disk}$ (see text for the
  definition of $\theta_{disk}$). The curves show a fit with a gaussian of the
  half-width $\Delta\theta_{disk}=18.5^\circ$.}
\label{gauss}
\end{figure}

The growth rate of the X-ray
flux between the phases $70^\circ<\theta<110^\circ$ is remarkably
close to the decrease rate at the phases $290^\circ<\theta<340^\circ$.
This enables us to make a conjecture that the observed decrease of the
X-ray flux can be associated to the exit of the pulsar from the disk.
To test this conjecture we superimpose the pre-periastron X-ray and
TeV lightcurves over the post-periastron lightcurves by shifting the
phase of the post-periastron data points by $-180^\circ$. For X-rays
we have used only data from the rise ($\theta<110^\circ$) and the fall
 ($\theta>290^\circ$) periods. The result
is shown in Fig. \ref{gauss}. One can see that in such representation
the rise and decrease of both X-ray and TeV flux from the system can
be well fitted with a gaussian curve
$F(\theta)\sim\exp\left(-(\theta-\theta_{disk})^2/(2\Delta\theta_{disk}^2)\right)$.
We find that the best fit is achieved with the parameter choice
$\theta_{disk}\simeq 109.1^\circ$, $\Delta\theta_{disk}\simeq
18.5^\circ$ (the coordinate $\Delta\theta$ along the X-axis of
Fig. \ref{gauss} is, in fact $\Delta\theta=\theta-\theta_{disk}$).

The most obvious explanation to the shape of the X-ray and TeV
lightcurves shown in Fig. \ref{gauss} is that pulsar crosses the thick
disk with the half opening angle $\Delta\theta_{disk}$. The ``middle''
plane of the disk intersects the pulsar orbital plane along the line
$\theta=\left(109.1^\circ\bigcup 289.1^\circ\right)$. The intersection
of the disk with the above parameters with the pulsar orbital plane is
shown schematically in Fig. \ref{xmm_ellips} by a shaded area.
Because of the exponential profile, the disk has no fixed boundary.
Denser shading in Fig. \ref{xmm_ellips} corresponds to the one opening
angle of the disk, while lighter shading corresponds to two opening
angles. The disk is also shown by the same shading in the Fig.
\ref{Xradio}.  From this figure one can see that although the disk
appears quite symmetric in terms of the orbital phase $\theta$, it is
highly asymmetric in terms of the time $\tau$ which measures days from
the periastron passage. For example, during the second
(post-periastron) disk passage, the rise phase
($250^\circ<\theta<290^\circ$) of X-ray and TeV lightcurves takes just
some 10 days, while the decay ($290^\circ<\theta<340^\circ$) takes
about 100 days.

\subsection{Timing Analysis}

For each of the five PN event files in 0.2 -- 10 keV energy range with
a 6 ms time resolution we 
searched for a possible pulsation around the nominal
47.6~ms period using the epoch-folding technique ({\tt efsearch} 
tool from the {\tt XRONOS} package). 
From the individual observations  the
best $3-\sigma$ upper limit on pulsation at a period of
47.6~ms is $\sim$9\%, compatible with previous similar analysis made with
ASCA \citep{kaspi95,hirayama99}. In order to perform a deeper search for 
pulsations we combined neighboring XMM observations (namely X7-X10, 
see. Table 1). As the time span of these observations covers only
small fraction of the binary orbit we did not make
corrections to the binary motion of the neutron star. Usage of the combined
XMM data improves the upper limit to $\sim$2\%. 

Data of BeppoSAX/MECS (1.5-11 keV) observations also were searched for 
pulsations. For deeper search we combined observations S2-S5 (see Table 2).
Due to small time period of performed observations we did not 
make correction for binary motion of the neutron star. In spite of
considerably smaller effective area of MECS instruments of BeppoSAX 
observatory, the effective exposure time of observations of the source 
was much larger, which resulted in a comparable upper limit on X-ray 
pulsation fraction. We have not found
pulsations with 3-$\sigma$ upper limit at $\sim$2\%.

\subsection{Spectral Analysis \label{specan}}

\begin{figure}
\includegraphics[width=9cm,angle=0]{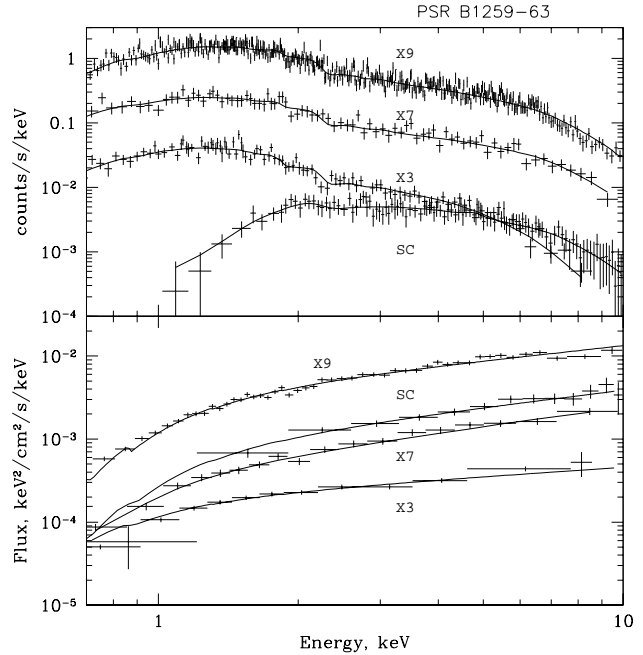}
\caption{Folded (upper panel) and unfolded (bottom panel) 
  \psrbsp spectra from the
  SC, X3, X7, and X9 observations. The \xmm\ spectra  for X7 and X9
  are for MOS1
  instrument. The X3 spectrum is for PN instrument.}
\label{spectry}
\end{figure}

The spectral analysis was done with NASA/GSFC XSPEC software package.
In Fig.~\ref{spectry} the folded and unfolded spectra of \psrb~ for SC, X3, X6,
and X9 observations are shown. 

\begin{figure*}
\begin{center}
\includegraphics[width=13cm,angle=0]{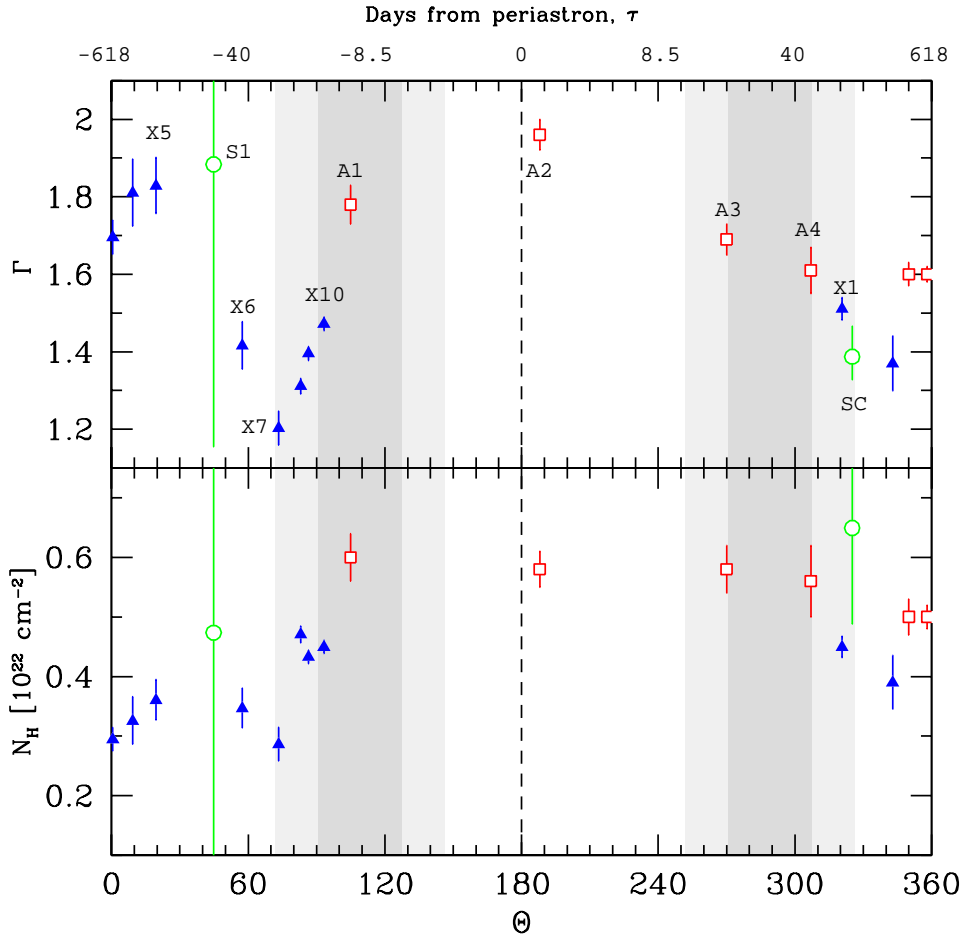}
\end{center}
\caption{Spectral parameters of  \xmm,\ \sax,   and 
\asca\ observations of the \psrb{} system. 
Notations are the same as in Fig. \ref{Xradio}. Top panel shows the
evolution of the  photon index $\Gamma$ as a function of the orbital
phase $\theta$.  Bottom panel shows the evolution of the hydrogen
column density $N_H$. Denser and lighter shaded region show phases
within one and two half opening angles of the Be star disk.}
\label{spechist}
\end{figure*}

A simple power law with a photoelectrical absorption describes the data well,
with no evidence for any line features. In Table \ref{summary}
we present the results of the three parameter fits to the \xmm\ data in
the 0.5 -- 10 keV energy range.
The uncertainties are given at the $1\sigma$
statistical level and do not include systematic uncertainties.
The graphical representation of the evolution of the \psrb\ spectral parameters
along the orbit is given in Figure \ref{spechist} which includes
data from the \sax\ and \xmm\ observations, and those from \textit{ASCA}
(taken from  \citet{hirayama99}).

\begin{table}
\caption{Spectral parameters for \sax\ and \xmm\ Observations of \psrb.$^*$\label{summary}}
\begin{center}
 \begin{tabular}{l|c|c|c|c}
\hline
Data  & $F$(1-10 keV) &$\Gamma$&$N_H$                &$\chi^2$ (dof) \\
Set   &$10^{-12}$erg s$^{-1}$& &($10^{22}$ cm$^{-2}$)& \\
\hline
S1& 1.36$^{+0.26}_{-0.43}$&1.88$^{+0.73}_{-0.22}$&0.47$^{+1.48}_{-0.47}$&0.814(201)\\
SC& 6.31$^{+0.69}_{-0.80}$&1.39$^{+0.06}_{-0.08}$&0.65$^{+0.16}_{-0.20}$&0.898(813)\\
X1& 8.51(30) &1.51(3)&0.45(2)&0.97 (341) \\ 
X2& 2.81(23) &1.36(7)&0.39(4)&1.02 (122) \\
X3& 1.04(49) &1.69(4)&0.29(2)&0.88 (207) \\
X4& 1.82(14) &1.82(7)&0.36(3)&1.01 (109) \\
X5& 0.99(9)  &1.80(8)&0.32(4)&0.80 ( 62) \\
X6& 2.47(17) &1.41(6)&0.35(3)&0.96 (151) \\ 
X7& 4.05(23) &1.20(4)&0.28(3)&0.95 (154) \\ 
X8& 12.59(34)&1.31(2)&0.47(1)&1.12 (614) \\
X9& 24.27(56)&1.39(2)&0.43(1)&1.07 (737) \\
X10&16.79(37)&1.47(2)&0.45(1)&0.88 (772) \\
\hline
\end{tabular}
\end{center}
$^*$ Number in parentheses represent 68\% confidence interval 
uncertainties in the last digit quoted.
\end{table}

\begin{figure*}
\begin{center}
\includegraphics[width=8.5cm,angle=0]{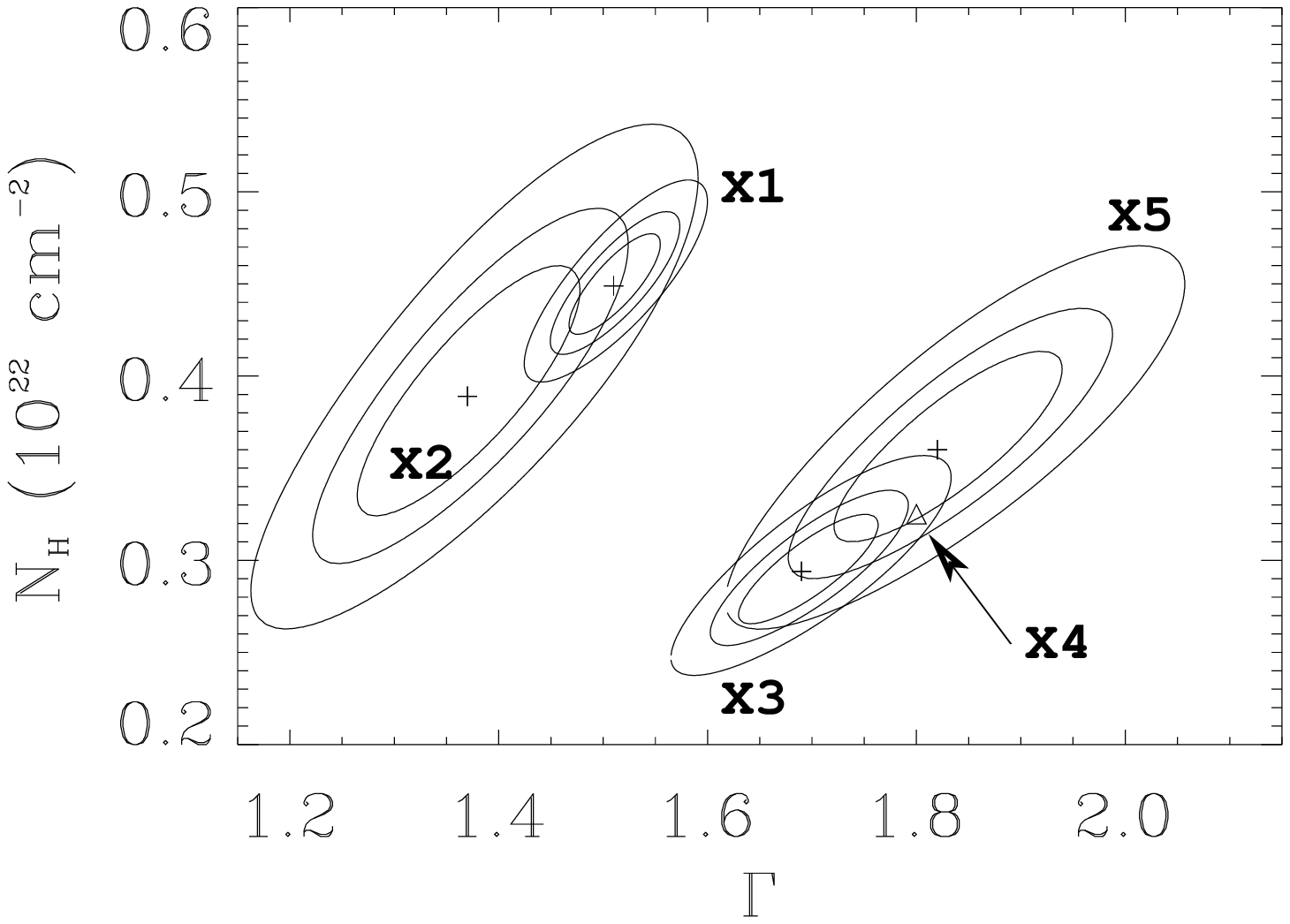}
\includegraphics[width=8.5cm,angle=0]{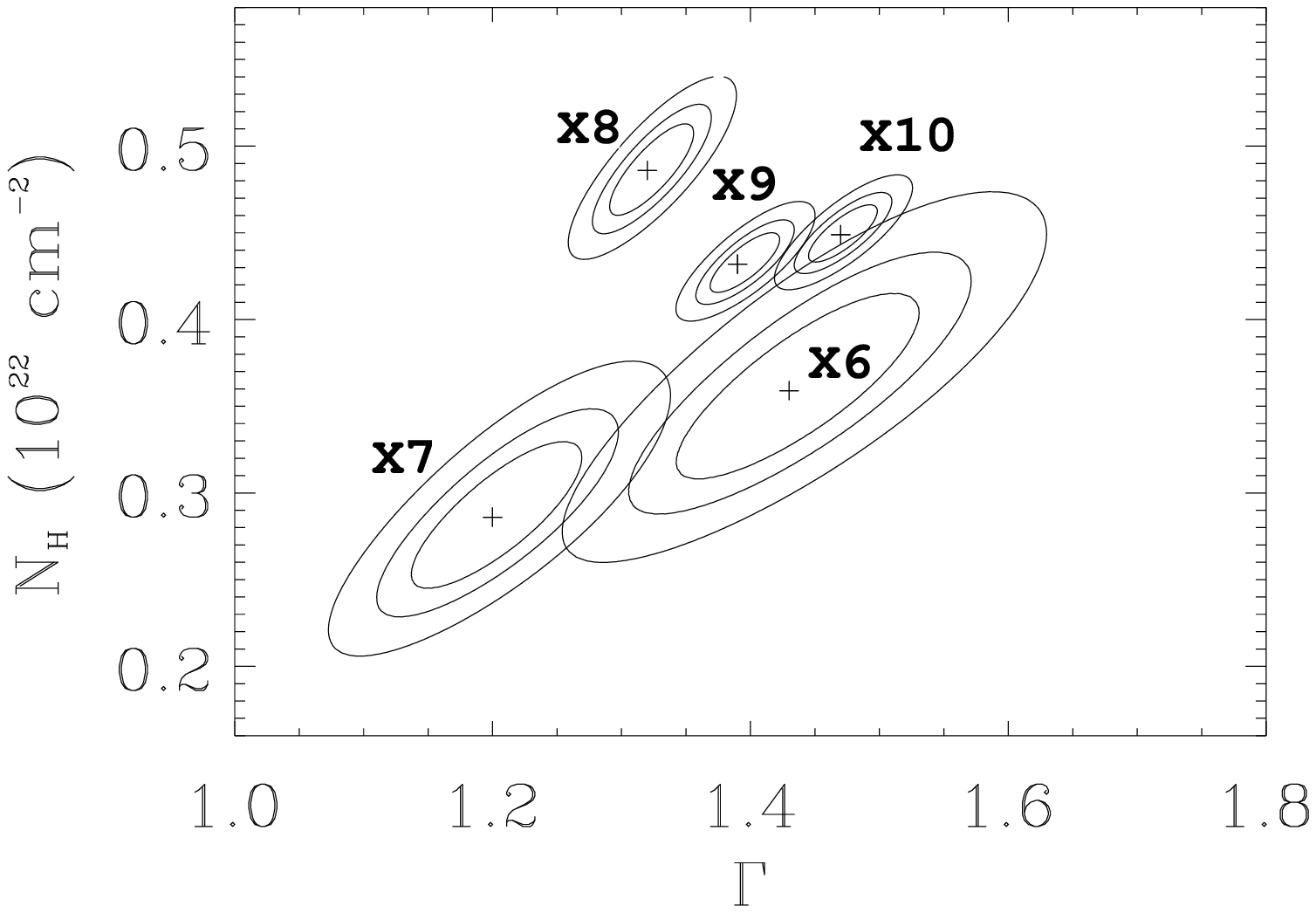}
\end{center}
\caption{Confidence contour plots of the column density $N_H$ and of the photon
spectral index $\Gamma$ uncertainties for a power-law fit to \xmmsp
observations. The contours give 68\%, 90\%, and 99\% confidence levels.
X4 observation, marked with triangle, is very similar to the X5 observation
and the contours are not plotted here.
\label{cntrspe}}
\end{figure*}

In Figure \ref{cntrspe} we show contour plots of $N_H$ versus photon
spectral index $\Gamma$, holding the 1 -- 10 keV flux fixed at the
values given in Table \ref{summary}. The contours plotted are the
68\%, 90\%, and 99\% confidence levels.  For the X5 observation the
values of $N_H$ and $\Gamma$ are consistent with X4 observation, and
we omitted the X4 contours to keep the Figure clear.
 
One can see that the variation of $N_H$ as a function of orbital phase
is correlated with the Be star disk crossing. No variations of $N_H$
are observed for the \xmm\ observations X2 to X7 which all fall
outside the disk (assuming the disk parameters found in 
Section 3.2). The mean value of of $N_H$ during these observations is
$N_H(X2-X7)\sim 0.32\times 10^{22}$cm$^{-2}$.  However, the
observations X8, X9, X10 and X1 are characterized by absorption which
is higher by a factor of about 1.5, $N_H(X8-X10,X1)\sim 0.45\times
10^{22}$cm$^{-2}$. Thus, the hydrogen column density remains high
starting from about 20 days before the periastron till at least 87
days after periastron.

The behaviour of the spectral slope as a function of the orbital phase
$\theta$ found in XMM observations differs from the one found in
\asca\ observations. Whereas in 1994 -- 1995 the spectrum at apastron
was found to be significantly harder than during periastron, \xmm\ 
observations show much softer spectrum at the
apastron with the spectral slope close to the \asca\ periastron value.  

The most remarkable feature of the spectral evolution of the system is
the hardening of the X-ray spectrum close to the moment when pulsar
enters the Be star disk at the phase $\theta\simeq
\theta_{disk}-2\Delta\theta_{disk}\simeq 70^\circ$. One can 
see that the decrease of the photon index $\Gamma$ is simultaneous
with the onset of the rapid growth of the X-ray flux.  To the best of our
knowledge, such behaviour was not predicted in any of existing models
of X-ray emission from the system.  Moreover, the observed values
$\Gamma<1.5$ during the observations X6 -- X10 are difficult to
reconcile with the models of synchrotron or inverse Compton emission
from shock-accelerated electrons because of the too hard electron
spectrum implied. Possible error in determination of $\Gamma$ due to
the uncertainty in determination of $N_H$ (see Fig. \ref{cntrspe}) is
too small to allow for $\Gamma\ge 1.5$ at least for observations X7 -- X9.

\section{Discussion}

In all theoretical models of unpulsed emission from \psrbsp the
collision of relativistic pulsar wind with non-relativistic wind of
the Be star plays a crucial role.  Due to the interaction of the winds
a system of two shock waves forms between the stars. Close to the Be
star and to the pulsar both winds are supposed to be radial but after
passing the shocks particles turn and start to flow along the contact
surface, losing their energy via synchrotron and inverse Compton (with
the seed photons being the Be star soft photons) emission.

In the model of \citet{cher99,cher00} (CI) the observed X-ray emission
is due to inverse Compton scattering of soft  photons from
the Be star on the same electron populaiton that produce the
non-pulsed radio emission from the system.  The intensity of observed
unpulsed X-ray and radio emission depends strongly on the flow
velocity of the relativistic particles beyond the shock. Slower drift
velocity leads to the higher concentration of the relativistic
electrons beyond the shock and higher X-ray and radio
luminosity. Decrease of the drift velocity during the pulsar passage
through the disk is caused by the macroscopical mixing of pulsar wind
with the stellar wind.

Main prediction of
this model is the correlation of X-ray with the radio lightcurves of the
system. Although different quality of X-ray and radio lightcurves does
not allow to make definitive conclusions about radio-X-ray
correlations, our \xmm\ observations show that the onset of rapid
increase of the X-ray flux is simultaneous with appearance of the
unpulsed radio emission from the system (see Fig \ref{Xradio}).
\begin{figure}
\includegraphics[width=\columnwidth,angle=0]{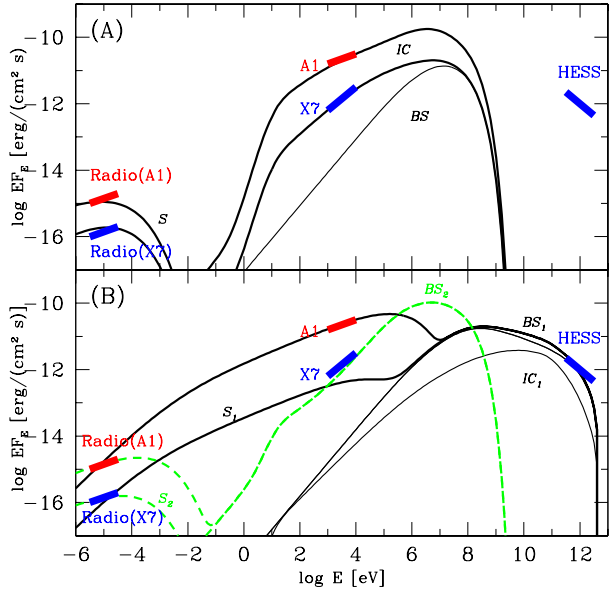}
\caption{Modification of the broad-band spectra  of
  CI (a) and TA (b) models due
  to the bremsstrahlung emission from shocked stellar/pulsar wind
  electrons during the disk passage. Contributions to the total spectrum
  from synchrotron, inverse Compton  and bremsstrahlung 
  processes are marked with S, IC and BS, respectively. In the case of
TA model black solid curves show the spectrum of emission from
the shocked pulsar wind while the green dashed curves show the
contribution from the shocked stellar wind. Parameters used for the
CI model calculation are
 $n=2\times 10^{9}$cm$^{-3}$, $B=0.01$~G
(low) and $B=0.1$~G (high synchrotron flux), $dN_e/d\gamma\sim
\gamma^{-2.4}\exp(-\gamma/400)$, with such parameters Coulomb break
  is at $\gamma\sim 50$. The same parameters are used for the
calculation of the shocked stellar wind emission (green dashed curves)
in the lower panel. The emission from the shocked pulsar wind was
calculated assuming  $n=2\times 10^{11}$cm$^{-3}$, $B=0.15$~G
(low) and $B=1.5$~G (high synchrotron flux), $dN_e/d\gamma\sim
\gamma^{-2.3}\exp(-\gamma/4\times 10^{6})$. With such parameters
 Coulomb break is at $\gamma\sim 500$.}
\label{models}
\end{figure}

The X-ray inverse Compton emission in the CI model is produced by
electrons with rather moderate gamma-factors, $\gamma_{e,IC}\sim 10$
while the gamma-factors of radio-synchrotron-emitting electrons are
around $\gamma_{e,S}\sim 10^2$. The synchrotron/inverse Compton
cooling times for such electrons are
\begin{eqnarray}
\label{synch}
t_{S}&\simeq& 10^8\left[\frac{0.1\mbox{
G}}{B}\right]^{3/2}\left[\frac{10\mbox{ GHz}}{\nu}\right]^{1/2}\mbox{
s}\nonumber\\ t_{IC}&\simeq& 10^6\left[\frac{1\mbox{
erg/cm}^3}{U_r}\right]
\left[\frac{1\mbox{ keV}}{\epsilon_{IC}}\right]^{1/2}\mbox{ s}
\end{eqnarray}
Both are large compared to the typical adiabatic cooling (or escape)
time $t_{esc}\sim R/\mbox{v}\sim 10^3$~s ($R$ is the characteristic size of
the emission region, and $\mbox{v}$ is the velocity of the flow along the
contact surface). Another important time scale for the electrons in
the energy range relevant in the CI model is the Coulomb loss scale
\begin{equation}
\label{coul}
t_{coulomb}\simeq 10^{4}\left[\frac{10^{9}\mbox{ cm}^3}{n}\right]\left[\frac{\gamma_e}{10}\right]\mbox{ s}
\end{equation}
As the pulsar enters the disk, the density of the stellar wind
increases. Typical estimates of the wind denisty at the pulsar
location (assuming the radial profile $n\sim (R_\star/R)^2$) result in
the density estimate $n\sim 10^9-10^{10}$~cm$^{-3}$. Thus, inside 
the disk Coulomb cooling time is comparable to the escape time.

This effect leads to the modification of electron spectrum at the
energies below the so-called ``Coulomb break'' energy at which
$t_{coulomb}=t_{esc}$:
\begin{equation}
\label{break}
\gamma_{e,coulomb}\simeq 1\left[\frac{n}{10^{9}\mbox{ cm}^{-3}}\right]\left[\frac{t_{esc}}{10^3\mbox{ s}}\right]
\end{equation}  
Namely, since the Coulomb loss rate $dE/dt\sim \gamma_e/t_{coulomb}$
is energy-independent, the electron acceleration spectrum with the
spectral index $\Gamma_e>2$ hardens  to $(\Gamma_e-1)<2$ below the break
energy. This naturally leads to the hardening of the spectrum of
inverse Compton X-ray emission at the moment when the pulsar enters
the disk, as it is shown in Fig. \ref{models}a. One can note that the
Coulomb break energy (\ref{break}) is close to the energy of the X-ray
emitting photons. This means that the X-ray photon index is very
sensitive to the slight changes of the density and/or of the outflow
velocity (see Fig. \ref{models}a in which the two spectra are
calculated for $\gamma_{e,coulomb}=10$ and
$\gamma_{e,coulomb}=50$). Since both vary during the disk passage (it
is assumed that within the disk pulsar and stellar
winds are  macroscopically  mixed), numerical modelling is required
 to predict the time evolution
of the X-ray spectral index. We leave the detailed modelling of this
for future work \citep{cher06}.

In the original CI model, only electorns with typical gamma-factors of
10-100 are considered. The presence of protons with comparable
gamma-factors $\gamma_p\sim 1-10^3$ at the contact surface would lead
to the generation of TeV emission during the disk passage which is due
to the proton-proton interactions in the disk (see
\citet{kawach04}). Macroscopic mixing of the pulsar and stellar wind will
lead to the increase of the proton-proton interaction rate in the
disk, which can explain the form of the TeV lightcurve of the system.

In the model of \citet{tavani97} (TA) the X-ray emission is the synchrotron
emission from  shock-accelerated electrons of the pulsar wind.
Typical gamma-factors of the X-ray emitting electrons are $\gamma\sim
10^6$. Modulation of the X-ray flux from the source during the disk
and periastron passage is due to the combined effect of displacement
of the contact surface by the disk (which leads to the variation of
the magnetic field strength) and the increase of the inverse Compton
energy loss close to the periastron.
 
One of the main predictions of TA model is
steepening of the X-ray spectrum during the disk and periastron
passage. The effect is due to the fact that synchrotron and/or inverse
Compton cooling time of X-ray emitting electron becomes comparable or
shorter than the time needed to escape from the emission region.
However, the observed behaviour of the spectrum is quite opposite: 
the entrance to the disc is accompanied by the hardening of the
spectrum. 

The values of the photon index $\Gamma<1.5$ found in the observations
X6 -- X10 are too hard to be explained with the synchrotron emission
from shock-accelerated electrons which normally result in the electron
spectra with power-law index $\Gamma_e\ge 2$ and, correspondingly in
synchrotron spectra with photon index $\Gamma\ge 1.5$. Of course, over
a large fraction of the pulsar orbit the synchrotron cooling time is
larger than the time needed to escape from the emission region along
the contact surface. Thus, if one assumes that the production spectrum
of high-energy electrons responsible for the X-ray emission is harder
than $\Gamma_e=2$, (if, e.g. electrons originate in the cold pulsar
wind with the bulk Lorentz factor $\gamma\sim 10^6$), it is possible,
in principle, to obtain spectra which are harder than $\Gamma=1.5$.
However, in this case one expects to observe the very hard X-ray
spectrum with $\Gamma<1.5$ over most of the pulsar orbit.

Within the TA model radio
synchrotron emission originates from different population of electrons
than the X-ray synchrotron emission (radio is convensionally
attributed to shock-accelerated stellar wind electrons, while X-ray
flux is emitted by the pulsar wind electrons).  Although it is not
possible to explain the hardening of the X-ray spectral index by the
changes in the pulsar wind electron spectrum, one can try to ascribe
the observed hardening to the contribution from the shocked stellar
wind electrons. The X-ray spectrum with photon index harder than
$\Gamma=1.5$ is natural if there is a bremsstrahlung contribution to
the X-ray flux. The increase of the density of stellar wind
inside the disk can lead to the increase of bremsstrahlung
luminosity.

The estimate of the bremsstrahlung loss time 
\begin{equation}
\label{brems}
t_{brems}=10^5\left[\frac{10^{10}\mbox{ cm}^{-3}}{n}\right]\mbox{ s}
\end{equation}
shows that for radio emitting electrons it is orders of magnitude
shorter than the synchrotron cooling time (see (\ref{synch})). Thus,
the bremsstrahlung luminosity should be much higher than the radio
synchrotron luminosity. From Fig. \ref{models} one can see that a
satisfactory fit to the broad-band spectrum can be obtained assuming
the disk density $n\sim 10^{11}$~cm$^{-3}$. It is interesting to note
that bremsstrahlung can also explain the behaviour of the TeV
lightcurve in this case, if one assumes the possibility of mixing of
the pulsar wind electrons with the pulsar disk medium (similar to the
one assumed in the CI model).

The bremsstrahlung model faces, however, a serious overall energetics
problem.  From Fig. \ref{models}b one can see that the total
bremsstrahlung luminosity is at least 2 orders of magnitude higher
than the X-ray luminosity at the moment of X7 observation,
$L_{X7}\simeq 10^{33}$~erg/s (see Fig. \ref{Xradio}). As it is
discussed above, the energy loss rate of mildly relativistic electrons
responsible for the bremsstrahlung emission is dominated by the
Coulomb losses. Comparing (\ref{brems}) to (\ref{coul}) one can see
that the Coulom loss rate is still 2-3 orders of magnitude higher than
the bremsstrahlung luminosity. Thus, to obtain the X-ray
bremsstrahlung flux at the level of $10^{33}$~erg/s one has to assume
that the power injected into the $10-100$~MeV electrons is about
$10^{38}$~erg/s which is much higher than the spin-down luminosity of
the pulsar $L_{pulsar}\simeq 10^{36}$~erg/s.

\section{Conclusions}

In this paper we have presented the \xmm\ observations of the pulsar
PSR B1259-63 during its 2004 periastron passage near the companion Be
star SS 2883. Combining our observations with the previous X-ray
observations from \xmm, \sax\ and \asca\ we  produced the
detailed lightcurve of the system over the wide range of the orbital
phases.

Using the large X-ray data sample we were able to constrain the
geometry of the Be star disk. In particular, we have found that the
half-opening angle of the disk (projected on the pulsar orbital plane)
is $\Delta\theta_{disk}\simeq 18.5^\circ$ and that the line of
intersection of the disk with the pulsar
orbital plane is inclined at about $70^\circ$ with respect to the
major axis of the pulsar orbit. 

Inspection of the X-ray and TeV lightcurves of the system has revealed
correlation of the X-ray and TeV flux variations. Our analysis shows that the
variations of the TeV flux some 10-100 days after the periastron
passage can be well fit with the gaussian curve whose parameters
coincide with the parameters of the disk found from the X-ray analysis
(center at $\theta_{disk}=289.1^\circ$, half-width
$\Delta\theta=18.5^\circ$) (see Fig. \ref{gauss}). If the observed
behaviour of the flux is indeed due to the influence of Be star
disk, it can not be explained within the inverse Compton scattering
model of TeV emission from the system.
 
Our \xmm\ observations which were done during the period of first
pulsar entrance to the disk (some 40 to 20 days before the periastron)
revealed quite unexpected evolution of the spectrum of X-ray emission.
In particular, the X-ray spectral slope hardened from $\Gamma\simeq
1.8$ down to $\Gamma\simeq 1.2$ roughly at the onset of the rapid
growth of the X-ray luminosity. This behaviour can not be explained
within the synchrotron model of X-ray emission from the system. We
have shown that the observed hardening can be the result of the
hardening of electron spectrum due to the Coulomb losses in the disk
in the model where X-ray emission is the inverse Compton emission from
the population of electrons responsible for the radio synchrotron
emission. Otherwise, the hardening can be caused by the increase of
the bremsstrahlung emission from these electrons. However, the
bremsstrahlung problem faces a serious energy balance problem.

\section{Acknowledgments}

The authors acknowledge useful discussions with F.~Aharonian and V. Beskin. 
Authors are grateful to V.~Kaspi  for helpful comments.
We are grateful to L.~Foschini for helpful advices on the \xmm\
data analysis and thank M.~Revnivtsev for the help with the \sax\ data
reduction. MC thank M. T\"urler for the help with figure production.
AL acknowledges the support of RFFI grant 04$-$02$-$17276.

 \label{lastpage}

\end{document}